\documentclass[twocolumn,superscriptaddress]{revtex4}

\usepackage{graphicx}
\usepackage{dcolumn}
\usepackage{bm}
\usepackage{physics}
\usepackage{amsmath}
\usepackage{ulem}

\newcommand{\bonnpi}{Physikalisches Institut, University of Bonn, Nussallee 12, 53115 Bonn, Germany}
\newcommand{\kl}{Physics Department and Research Center OPTIMAS, Technische Universit\"at Kaiserslautern, 67663 Kaiserslautern, Germany.}

\usepackage{tikz}
\usetikzlibrary{calc}
\usetikzlibrary{shapes}
\usepackage{ifthen}
\usepackage{booktabs}
\usepackage{multirow}

\newcommand{\upd}[1]{^\mathrm{#1}}
\newcommand{\ind}[1]{_\mathrm{#1}}

\begin{document}

\title{Floquet-engineered pair and single particle filter in the Fermi Hubbard model}

\date{\today}

\begin{abstract}
We investigate the Fermi-Hubbard model with a Floquet-driven impurity in the form of a local time-oscillating potential. For strong attractive interactions a stable formation of pairs is observed. These pairs show a completely different transmission behavior than the transmission that is observed for the single unpaired particles. Whereas in the high frequency limit the single particles show a maximum of the transition at low driving amplitudes, the pairs display a pronounced maximum transmission when the amplitude of the driving lies close to the ratio of the interaction $U$ and the driving frequency $\omega$. We use the distinct transmission behaviour to design filters for pairs or single particles, respectively. For example one can totally block the transmission of single particles through the driven impurity and allow only for the transmission of pairs. We quantify the quality of the designed filters.
\end{abstract}
\author{Friedrich H\"ubner}
\affiliation{\bonnpi}
\author{Christoph Dauer}
\affiliation{\kl}
\author{Sebastian Eggert}
\affiliation{\kl}
\author{Corinna Kollath }
\affiliation{\bonnpi}
\author{Ameneh Sheikhan}
\affiliation{\bonnpi}

\maketitle
By the extended demand for the miniaturization of technical devices as transistors, nowadays a large effort is made in order to engineer quantum devices which work at the few particle level. Whereas in earlier studies mainly steady state working principles have been investigated, during the last decade also the dynamic shaping and controlling of such devices attracted an increasing attention. Progress in this field is rewarding since it opens the path to using the time-dependent complexity as a resource for novel states and tuning possibilities. However, the microscopic understanding of non-equilibrium situations of these quantum systems in particular in the presence of interaction remains a huge challenge. 

One interesting and promising research direction in the last decade is dynamic driving of quantum systems with time-periodic fields in order to control their characteristic behavior~\cite{Eckardt_2015}. A periodic drive has been applied in order to obtain the dynamical localization~\cite{PhysRevLett.99.220403}, artificial magnetic fields~\cite{PhysRevLett.107.255301,KollathBrennecke2016, SheikhanKollath2016, SheikhanKollath2016R, SheikhanKollath2019} to phase transitions~\cite{PhysRevLett.102.100403,PolettiKollath2011, PhysRevB.94.174503,PhysRevLett.126.243401,PhysRevResearch.2.013275,SheikhanKollath2020} or to control the bound pairs \cite{PhysRevA.80.063409}.

Theoretically, periodically modulated systems can be described using the 
so-called Floquet theory\cite{Eckardt_2015}. The time-periodic symmetry allows to solve the time-dependent Schr\"odinger equation in terms
of an eigenvalue problem with a conserved quasi-energy, analogous to the quasi-momentum in 
Bloch's theorem.  The Floquet solution corresponds to a stable steady state of the 
system and also results in an effective Floquet Hamiltonian for the stroboscopic time evolution.

The topic of this paper is the interplay of a periodically driven impurity and interactions in quantum systems. Already on the single particle level, some very interesting results for the transmission of time-periodic {\it local} fields have been obtained.  One example is a periodically modulated quantum dot or quantum point contact which has been predicted to control the transmission of single particles through the dot \cite{reyes2017transport,PhysRevB.93.180301} or through the quantum point contact \cite{PhysRevB.103.L041405}.  Further, a periodically modulated quantum dot has been proposed to induce bound states \cite{PhysRevB.96.104309} or as a spin filter~\cite{PhysRevLett.119.267701} for which one spin species is fully blocked while the other is fully transmitted by the dot. 

As we show here, the effect of the interaction in combination with a time-periodic impurity field enhances the complexity considerably. In particular, for effective attractive interactions between fermionic particles as they occur in superconductors, we find that the formation of interaction induced pairs leads to a highly non-trivial transmission behavior as a function of driving amplitude and frequency, which is very distinct from the single particle behavior.  We show that the combination of interaction and non-equilibrium driving can indeed be used as a resource to construct novel devices such as a pair filter or pair blocker.

We consider a Fermi-Hubbard chain with a driven chemical potential at site $0$ described by the Hamiltonian
\begin{align}
\vb{H}(t) = -J&\sum_{n,\sigma} \qty(\vb{c}_{n\sigma}^\dagger\vb{c}_{n+1\sigma} + \mathrm{h.c.}) + U\sum_n \vb{n}_{n\uparrow}\vb{n}_{n\downarrow}\nonumber\\
 &+ \lambda\omega\cos(\omega t) \qty(\vb{n}_{0\uparrow} + \vb{n}_{0\downarrow}).
\label{equ:hamiltonian}
\end{align}
Here the operator $\vb{c}^{(\dagger)}_{n\sigma}$ annihilates (creates) a fermion with spin $\sigma \in \qty{\uparrow, \downarrow}$ on site $n$, $\vb{n}_{n\sigma} = \vb{c}^\dagger_{n\sigma}\vb{c}_{n\sigma}$ is the density operator, $J$ is the hopping parameter, $U < 0$ the attractive Hubbard  interaction, and  $\omega$ and $\lambda$ are the frequency and unit less amplitude of the driving, respectively. Here, we set $\hbar=1$ and measure the lengths in units of the lattice spacing. We will investigate the scattering of a single incoming particle or of an incoming on-site pair of spin up and spin down fermions at the impurity. 

\begin{figure}[hbtp]
\centering
\scalebox{0.4}{
	\begin{tikzpicture}[scale=1]
	\tikzstyle{every node}=[font=\huge]
	\def\size{4}
	\def\spacing{1.8}
	\def\spacingy{2}
	\def\circlesize{0.3}
	\def\linelength{0.3}
	\def\linewidth{3}

	\def\subplotsingle{-1.7*\spacingy}
	\def\subplotpair{-3.2*\spacingy}

	\def\cfloquet{blue}	
	
	
	\foreach \x in {-\size,...,-1} {	
		\coordinate (lattice\x) at (\x*\spacing,0);
	}
	\coordinate (lattice0) at (0,0.3*\spacing);
	\coordinate (lattice0s1) at (0,0.0*\spacing);
	\coordinate (lattice0s2) at (0,-0.3*\spacing);
	\foreach \x in {1,...,\size} {	
		\coordinate (lattice\x) at (\x*\spacing,0);
	}
	\filldraw[color=lightgray] (lattice0s2) circle (\circlesize);
	\draw[line width=\linewidth,color=lightgray] (lattice-1)--(lattice0s2);
	\draw[line width=\linewidth,color=lightgray] (lattice1)--(lattice0s2);
	\filldraw[color=gray] (lattice0s1) circle (\circlesize);
	\draw[line width=\linewidth,color=gray] (lattice-1)--(lattice0s1);
	\draw[line width=\linewidth,color=gray] (lattice1)--(lattice0s1);
	\foreach \x in {-\size,...,\size} {
		\filldraw (lattice\x) circle (\circlesize);
	}
	\foreach \x in {-\size,...,\size} {
		\node at (lattice\x) [above=10] {\x};
	}
	\coordinate (lattice\number\numexpr \size + 1 \relax) at ({(\size+\linelength)*\spacing},0);
	\coordinate (lattice\number\numexpr -\size - 1 \relax) at ({-(\size+\linelength)*\spacing},0);  
	\foreach \x [evaluate=\x as \xp1 using \x+1]in {\number\numexpr -\size-1\relax,...,\size} {
		\draw[line width=\linewidth] (lattice\x)--(lattice\xp1);
	}
	\foreach \x  in {\number\numexpr \size-1\relax,...,\number\numexpr \size-1\relax} {
		\node () at ({(\x+0.5)*\spacing},-0.5) {$J$};
	}
	
	\draw[line width=3,<->,color=\cfloquet] (0.5,-1)--(0.5,1);
	

	\draw[dashed,line width=1] (-\size*\spacing-0.5*\spacing,0.5*\subplotsingle)--(\size*\spacing+0.5*\spacing,0.5*\subplotsingle);			

	\node at (-\size*\spacing, \spacing) [anchor=east] {(a)};	
	
	\def\lsfloquet{loosely dashed}
	
	\foreach \x in {-\size,...,\size} {
		\coordinate (singlelattice\x) at (\x*\spacing,\subplotsingle);
	}
	\foreach \x in {-\size,...,\size} {
		\filldraw (singlelattice\x) circle (\circlesize);
	}
	\coordinate (singlelattice\number\numexpr \size + 1 \relax) at ({(\size+\linelength)*\spacing},\subplotsingle);
	\coordinate (singlelattice\number\numexpr -\size - 1 \relax) at ({-(\size+\linelength)*\spacing},\subplotsingle);  
	
	\foreach \x [evaluate=\x as \xp1 using \x+1]in {\number\numexpr -\size-1\relax,...,\size} {
		\def\tmplinestyle{solid}
		\ifthenelse{\equal{\x}{-1}}{\def\tmplinestyle{\lsfloquet}}{}
		\ifthenelse{\equal{\x}{0}}{\def\tmplinestyle{\lsfloquet}}{}
		\draw[\tmplinestyle,line width=\linewidth] (singlelattice\x)--(singlelattice\xp1);
	}
	\foreach \x  in {-\size,...,\number\numexpr \size-1\relax} {
		\def\tmplabel{}
		\ifthenelse{\equal{\x}{-1}}{\def\tmplabel{$J \gamma\ind{s}$}}{}
		\ifthenelse{\equal{\x}{0}}{\def\tmplabel{$J \gamma\ind{s}$}}{}
		\ifthenelse{\equal{\x}{\number\numexpr \size-1\relax}}{\def\tmplabel{$J$}}{}
		\node () at ({(\x+0.5)*\spacing},\subplotsingle-0.5) {\tmplabel};
	}
	
	\node at (-\size*\spacing,\subplotsingle+0.5*\spacing) [anchor=west] {Single particle};
	\node at (-\size*\spacing,\subplotsingle+0.5*\spacing) [anchor=east] {(b)};
	
	\foreach \x in {-\size,...,\size} {
		\def\y{0}		
		\ifthenelse{\equal{\x}{-1}}{\def\y{0.5*\spacing}}{}
		\ifthenelse{\equal{\x}{0}}{\def\y{1.0*\spacing}}{}
		\ifthenelse{\equal{\x}{1}}{\def\y{0.5*\spacing}}{}
		\coordinate (pairlattice\x) at (\x*\spacing,\subplotpair+\y);
	}
	\foreach \x in {-\size,...,\size} {
		\filldraw (pairlattice\x) circle (\circlesize);
	}
	\coordinate (pairlattice\number\numexpr \size + 1 \relax) at ({(\size+\linelength)*\spacing},\subplotpair);
	\coordinate (pairlattice\number\numexpr -\size - 1 \relax) at ({-(\size+\linelength)*\spacing},\subplotpair);  
	
	\draw[dotted,line width=1] (pairlattice-\size)--(pairlattice\size);		
	
	\foreach \x [evaluate=\x as \xp1 using \x+1]in {\number\numexpr -\size-1\relax,...,\size} {
		\def\tmplinestyle{solid}
		\ifthenelse{\equal{\x}{-1}}{\def\tmplinestyle{\lsfloquet}}{}
		\ifthenelse{\equal{\x}{0}}{\def\tmplinestyle{\lsfloquet}}{}
		\draw[\tmplinestyle,line width=\linewidth] (pairlattice\x)--(pairlattice\xp1);
	}
	\foreach \x  in {-\size,...,\number\numexpr \size-1\relax} {
		\def\tmplabel{}
		\ifthenelse{\equal{\x}{-1}}{\def\tmplabel{\fontsize{19}{10}$J\ind{p}\gamma\ind{p} $}}{}
		\ifthenelse{\equal{\x}{0}}{\def\tmplabel{\fontsize{19}{10}$J\ind{p}\gamma\ind{p} $}}{}
		\ifthenelse{\equal{\x}{\number\numexpr \size-1\relax}}{\def\tmplabel{$J\ind{p}$}}{}
		\node () at ({(\x+0.5)*\spacing},\subplotpair-0.5) {\tmplabel};
	}

	\draw[line width=1,<->,color=black] ($(0,\subplotpair)!0.05!(pairlattice0)$) -- ($(0,\subplotpair)!0.75!(pairlattice0)$);
	\draw[line width=1,<->,color=black] ($(-\spacing,\subplotpair)!0.05!(pairlattice-1)$) -- ($(-\spacing,\subplotpair)!0.60!(pairlattice-1)$);
	\draw[line width=1,<->,color=black] ($(\spacing,\subplotpair)!0.05!(pairlattice1)$) -- ($(\spacing,\subplotpair)!0.60!(pairlattice1)$);
	
	\node () at (0.48*\spacing,\subplotpair+0.3*\spacing) {\fontsize{19}{10}{$2J\ind{p}\mu\ind{p}$}};
	\node () at (-\spacing+0.5*\spacing,\subplotpair+0.2*\spacing) {\fontsize{19}{10}{$J\ind{p}\mu\ind{p}$}};
	
	\node at (-\size*\spacing,\subplotpair+\spacing) [anchor=west] {Pair};
	\node at (-\size*\spacing,\subplotpair+\spacing) [anchor=east] {(c)};

\end{tikzpicture}
}
\caption{(a) Sketch of the Fermi-Hubbard chain with a periodically driven potential at site 0. (b,c) Effective high frequency models describing the scattering of a single particle (b) and a pair (c), respectively. For the single particle and the pair the hopping to and from the impurity site (dashed line) is reduced by a factor $\gamma\ind{s,p}$, respectively. For the pair, additionally, a triangular potential at sites $-1,0,1$ arises.}
\label{fig:setup}
\end{figure}
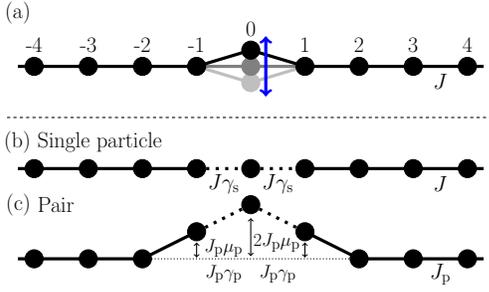
 
The scattering of a single particle at the periodically driven impurity has been studied in detail in Refs.~\cite{PhysRevB.93.180301}. 
At high driving frequencies $\omega \gg J$, an effective time-independent Floquet Hamiltonian can be derived using the high-frequency expansion~\cite{doi:10.1063/1.4916324,Eckardt_2015}. Such effective model relates to the time-evolution of the original model after full periods of the driving \footnote{The effective Hamiltonian does not cover the micromotion.}.
For a single particle impinging on the periodically modulated impurity  the effective Hamiltonian is given by
\begin{align}
\vb{H}\upd{s} = &-J\sum_{n\neq -1,0} (\vb{c}_{n\uparrow}^\dagger\vb{c}_{n+1\uparrow} + \mathrm{h.c.})\nonumber \\
&- JJ_0(\lambda)\sum_{n = -1,0} (\vb{c}_{n\uparrow}^\dagger\vb{c}_{n+1\uparrow} + \mathrm{h.c.})
\label{equ:hamiltonian_single}
\end{align}

\begin{figure}[hbtp]
\centering
\includegraphics[width=.35\textwidth]{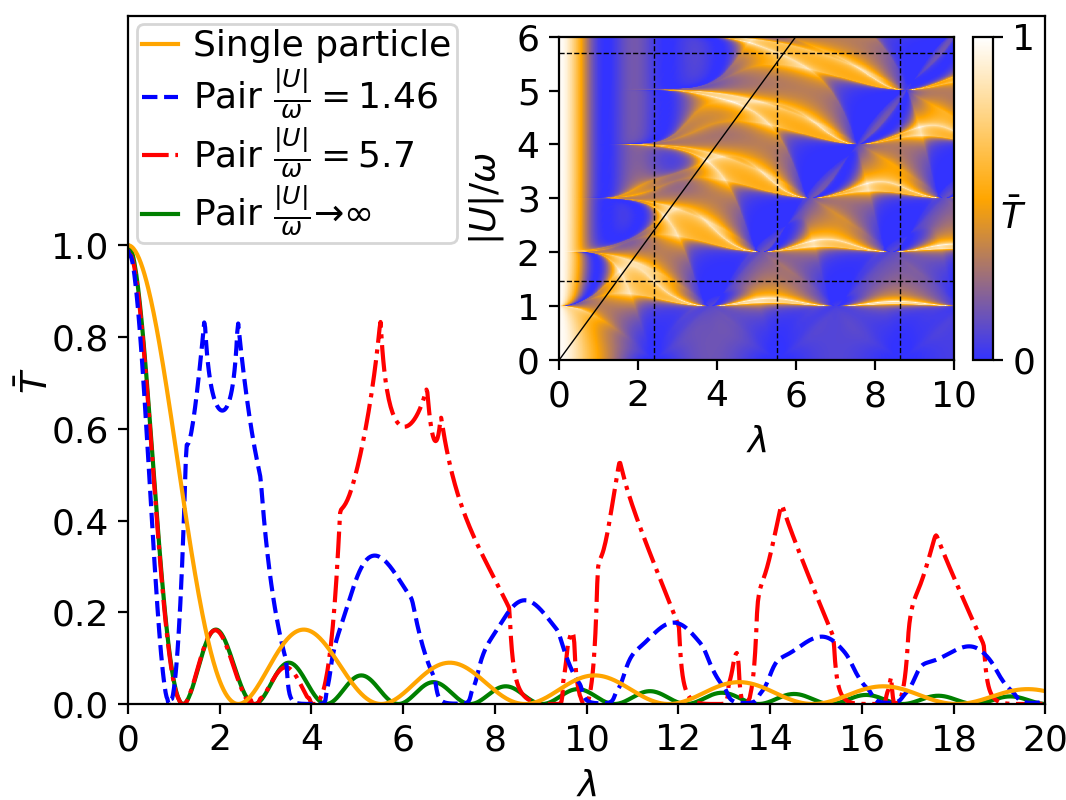}
\caption{Transmission in the limit $J \ll \omega,|U|$ as a function of the driving amplitude $\lambda$ for a single particle (yellow line) and a pair at $|U|/\omega = 1.46$ (blue line), $|U|/\omega = 5.7$ (orange line), $|U|/\omega \to \infty$ (green line). The inset shows the pair transmission versus $\lambda$ and $|U|/\omega$. The vertical lines show the values of $\lambda$ where the single particle transmission vanishes. The horizontal lines show the values of $|U|/\omega$ which are shown in the main panel. The solid line shows $\lambda = |U|/\omega$ which is roughly the position of the first maximum.}
\label{fig:transmission}
\end{figure}

This effective model (see Fig. \ref{fig:setup}) describes a single particle on a chain with a reduced hopping amplitude from and to the impurity site $0$. The effective hopping amplitude is reduced by the factor $\gamma\ind{s} = J_0(\lambda)$ which strongly depends on the driving amplitude $\lambda$. 
For a single particle with momentum $k$ (in units of one over lattice spacing) the transmission through the impurity becomes
$T^s_k = \left (1 + \qty(\frac{1}{\gamma\ind{s}^2}-1)^2\cot^2 k\right )^{-1}$ \cite{reyes2017transport}.
Further, the momentum averaged transmission defined as $\bar{T} = \int_{-\pi}^\pi T_k \frac{\dd{k}}{2\pi}$, one can explicitly compute for a single particle  giving
  $\bar{T}\ind{s} 
  = \gamma\ind{s}^2 = J_0(\lambda)^2$.
This means that the transmission of a single particle (see Fig. \ref{fig:transmission}) has a maximum at $\lambda=0$, then strongly decays to its first zero at $\lambda \approx 2.4$.  
For larger values of $\lambda$ an oscillating behaviour is seen with a decaying amplitude. Therefore, the transmission of a single particle can be regulated by the amplitude of the driving.
The analysis is not just limited to high frequencies in the single particle case. Interesting resonances have been found at low driving frequencies in the momentum resolved transmission using the full analysis of the
non-interacting problem in Ref.~\cite{reyes2017transport}. The integrated transmission can be obtained in the adiabatic limit as an expansion for small $\lambda \omega$ to lowest order as 
$\bar{T}\ind{s}^{\textrm{low}} \approx 1- \frac{\lambda\omega}{\pi J}$. 

Let us now consider the intriguing effect of the interaction $U$.  For strong attraction 
$-U \gg J, \omega$ using a Schrieffer-Wolff transformation the problem maps onto a model of stable entangled pairs which behave as non-interacting composite particles
\begin{align}
\vb{H}\upd{eff} = -J\ind{p}\sum_{n} \qty(\bm{\eta}_n^+ \bm{\eta}_{n+1}^- + \mathrm{h.c.}) + 2\lambda \omega\cos(\omega t) \bm{\eta}_0^+\bm{\eta}_0^-,
\nonumber
\end{align}
where we defined the pair creation/annihilation operators 
$\bm{\eta}_n^+ = \vb{c}_{n\uparrow}^\dagger\vb{c}_{n\downarrow}^\dagger$ and $\bm{\eta}_n^- = \vb{c}_{n\downarrow}\vb{c}_{n\uparrow}$ and the effective pair hopping parameter $J\ind{p} = \frac{2J^2}{|U|}$. This effective Hamiltonian $\vb{H}\upd{eff}$ clearly resembles the initial one Eq.~\ref{equ:hamiltonian} besides an effective tunneling amplitude $J_p$ and driving amplitude $2\lambda$ of the impurity.
Therefore, the scattering properties in the strong interacting limit can be derived from the known behaviour for single particles. For example  in the large frequency limit the pair transmission becomes $\bar{T}_{\rm P} \approx J_0(2 \lambda )^2$. Typically, due to the rescaled values, the single particle transmission can still be evaluated in the low frequency limit $\lambda \omega \ll J $ and $J<\omega$ where $\bar{T}\ind{s}^{\textrm{low}}$ 
is valid, whereas for the pair transmission already the described high frequency limit needs to be taken. This implies that if the limit of large interaction is taken first, the pair transmission is typically much smaller than the single particle transmission. 

The most interesting and also most complicated physical behavior occurs when the frequency is comparable to the interaction strength \footnote{ If $U$ is close to an integer multiple of $\omega$ the driving can break the pair into two single particles. A first approximation gives that in order to avoid pair breaking, that the distance of $U/\omega$ to the nearest integer should be below $4 J/\omega$. Note, that this implies that the maximum possible value of $J/\omega$ is $J/\omega = 0.125$. }.
In order to tackle this case, we consider 
the limit $J \ll \omega, |U|$.  It is indeed possible to have two leading energy scales and
make a single rotation by using a
Floquet-Schrieffer-Wolff transformation~\cite{PhysRevLett.116.125301}
\begin{align}
\vb{H}\upd{p} = -J\ind{p}\Bigl[&\sum_{n\neq -1,0} \qty(\bm{\eta}_n^+ \bm{\eta}_{n+1}^- + \mathrm{h.c.})\nonumber\\ 
&+ \gamma\ind{p} \sum_{n = -1,0} \qty(\bm{\eta}_n^+ \bm{\eta}_{n+1}^- + \mathrm{h.c.})\nonumber\\ &+ \mu\ind{p} (\bm{\eta}_{-1}^+ \bm{\eta}_{-1}^- + 2\bm{\eta}_{0}^+ \bm{\eta}_{0}^- + \bm{\eta}_{1}^+ \bm{\eta}_{1}^-)\Bigr].
\label{equ:hamiltonian_pair}
\end{align}
The periodic modulation of the impurity leads to the effective hopping amplitude and potential for the pair
\begin{align}
\gamma\ind{p} = \sum_l \frac{\frac{|U|}{\omega} (-1)^l J_l(\lambda)^2}{\frac{|U|}{\omega}-l}\; \textrm{and}\;  
\mu\ind{p} = \sum_l \frac{\frac{|U|}{\omega} J_l(\lambda)^2}{\frac{|U|}{\omega}-l} - 1\nonumber
\end{align}
respectively. Thus, within the effective Floquet description, the pair is subjected to a scattering at a region of three sites with scaled tunneling amplitude and a triangular potential (see Fig. \ref{fig:setup}).\\


Using the effective model (\ref{equ:hamiltonian_pair}) we calculate in the appendix that the momentum dependent pair transmission through the impurity becomes:
\begin{align}
T^p_k &= \frac{\gamma^4}{1 + \qty(\frac{\varepsilon^p(k)}{\sin{k}})^2}\frac{1}{\qty[\varepsilon^p(k)^2-\gamma^2]^2 +\varepsilon^p(k)^2\sin^2{k}}
  \label{equ:transmission_pair}
\end{align}
with $\varepsilon^p(k)=\cos{k}-\mu_p$. We numerically integrate this expression in order to obtain the momentum averaged transmission $\bar{T}\ind{p}$ shown in Fig.~\ref{fig:transmission}. Its behaviour is much more intriguing than the single particle transmission and the overall shape shows several pronounced features. (i) Firstly, similarly to the single particle transmission for low driving amplitude the pair transmission is one (i.e.~the system is fully transparent) at $\lambda=0$ and decays quickly  with increasing $\lambda$. (ii) Secondly, a remarkable feature is a pronounced maximum close to $\lambda \propto U/\omega$. (iii) For $\lambda > |U|/\omega$ the averaged transmission shows oscillations with decaying amplitude, giving  alternating regions of high and low transmission. These features are crucial in order to design quantum filters and we give a more detailed analysis of the origin of these features in the following. 

(i) The regime of small $\lambda \ll |U|/\omega$ can be understood by taking the limit $|U|/\omega \to \infty$ for which we obtain (see~\cite{NIST:DLMF}~\S 10.23) which gives 
$\gamma\ind{p} \to \sum_l (-1)^l J_l(\lambda)^2 = J_0(2\lambda)$
and 
$\mu\ind{p} \to \sum_l J_l(\lambda)^2 - 1 = 0$. 
This agrees with the results obtained from the large interaction limit described by $\vb{H}\upd{eff}$ 
if additionally the large frequency limit is taken.
We plot this result for $|U|/\omega \to \infty$ in figure \ref{fig:transmission} in red. It well approximates the initial decay of the transmission versus $\lambda$ and provides a good approximation for $\lambda \ll |U|/\omega$.\\
(ii) For $\lambda \approx |U|/\omega$, a maximum occurs in the pair transmission. Due to the prefactor $\frac{1}{\frac{|U|}{\omega}-l}$ the expression for $\gamma\ind{p}$ and $\mu\ind{p}$ are dominated by the Bessel function $J_{l_\pm}(\lambda)$ with the $l_\pm$ being the two integers closest to $\frac{|U|}{\omega}$. Both Bessel functions have their maximum around $\lambda \approx l_{\pm}$ which leads to the observed maximum of the transmission. Since we can tune the position of the maximum by tuning the value of the interaction, we will use it in order to design pair filters.  \\
(iii) 
For $\lambda > |U|/\omega$ the transmission exhibits alternating regions of high and low transmission.  The oscillations of the alternating regions roughly correspond to the oscillations of $J_{l_\pm}(\lambda)$. The detailed description of the amplitude of the transmission, however, requires taking into account more contributions. As $\lambda$ increases the averaged pair transmission also decreases and vanishes in the limit $\lambda \to \infty$ since $\gamma\ind{p} \to 0$.

In the following we will use our gained understanding on the features of the single particle and pair transmissions in order to design filters. Here we concentrate on the momentum integrated transmission.  

\begin{figure}[hbtp]
\centering
\includegraphics[width=.35\textwidth]{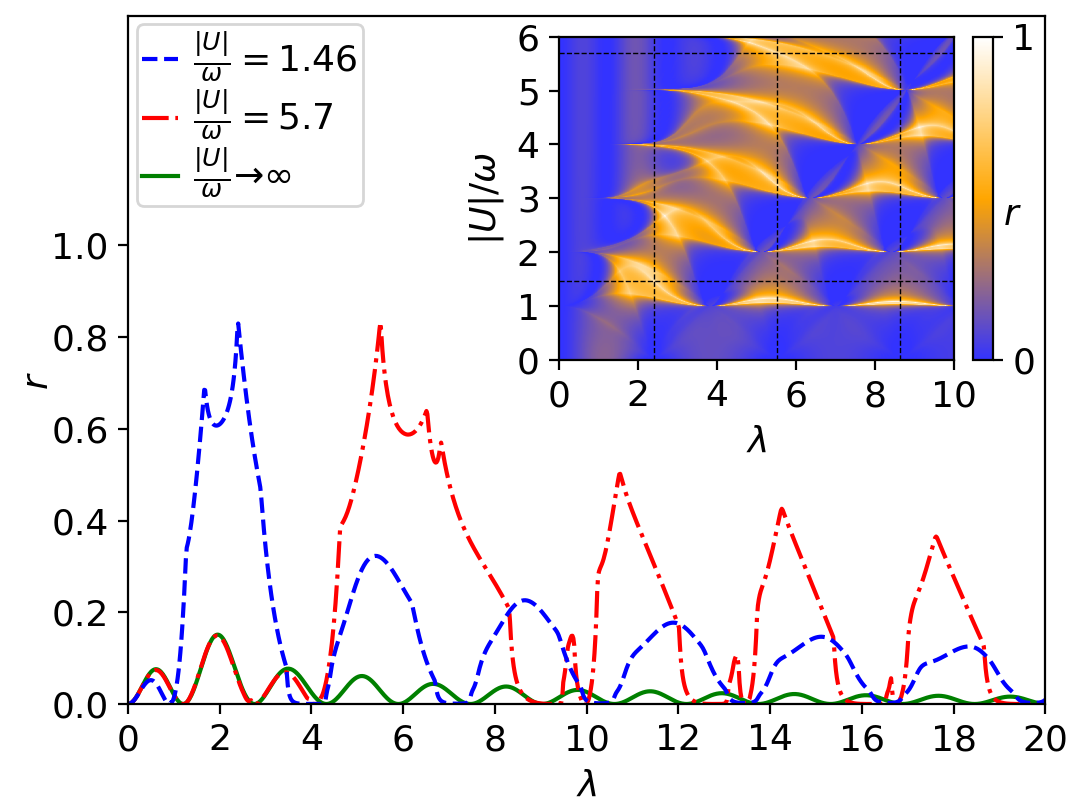}
\caption{Quality measurement for the pair filter, $r = \bar{T}\ind{p}\qty(1-\bar{T}\ind{s})$ versus the driving amplitude $\lambda$ for $|U|/\omega = 1.46$ (blue), $|U|/\omega = 5.7$ (orange) and $|U|/\omega \to \infty$ (green). $r$ gives the probability that a pair is transmitted while a single particle is reflected. The inset shows $r$ as function of $\lambda$ and $|U|/\omega$. The vertical lines show the values of $\lambda$ where the single particle transmission vanishes. The horizontal lines show the values of $|U|/\omega$ which are shown in the main panel.
 }
\label{fig:pair}
\end{figure}

We start to design a pair filter, which should mainly transmit pairs  through the impurity. In order to quantify the quality of such a filter, we consider the quantity 
$r = \bar{T}\ind{p}\qty(1-\bar{T}\ind{s})$.
This measures the product of the transmission of pairs and the reflection of single particles at the impurity. Therefore, $r$ gives the probability that the pair is transmitted while the single particle is reflected. The maximum value of $r$ is $r = 1$ which corresponds to a perfect pair filter, i.e.~only pairs can cross the impurity and these are transmitted with a probability equal one. The quantity $r$ is plotted in Fig.~\ref{fig:pair} using the same values of $|U|/\omega$ as in Fig.~\ref{fig:transmission}.

For $\lambda \gtrsim 2$, $r$ shows similarities to the pair transmission $\bar{T}\ind{p}$. In particular, the maxima for $\lambda \sim U/\omega$ and the oscillating structure of the additional local maxima persist.  This behaviour has its origin in the fact that for $\lambda > 2$ the single particle transmission is small such that one can approximate $r \approx \bar{T}\ind{p}$. However, for $\lambda\lesssim 2$ the single particle transmission typically gives an important contribution and reduces drastically the value of $r$. Therefore, the maxima of $r$ for $\lambda \sim U/\omega$ for $\lambda \gtrsim 2$ are typically the optimal values for a pair filter. They often reach above the value of $r\gtrsim 0.8$ which constitutes already a good pair filter. If one requires additionally that the single particle is fully blocked, the most prominent values of $\lambda$ for a pair filter are those were the single particle transmission vanishes. This happens at the zeros of the Bessel function $J_0(\lambda)$, e.g.~ $\lambda \approx 2.4, 5.52, 8.65, \ldots$. We mark these values in the inset of Figs.~\ref{fig:transmission} and \ref{fig:pair} as vertical lines. The values $|U|/\omega$ which we chose for the plots are optimal values for the first two zeros $\lambda \approx 2.4$ and $\lambda \approx 5.52$. For $|U|/\omega = 1.46$ the averaged pair transmission at $\lambda = 2.4$ is $\bar{T}\ind{p} = 0.79$ while for $|U|/\omega = 5.7$ at $\lambda = 5.52$ it is $\bar{T}\ind{p} = 0.81$. This means that while the single particles are fully blocked also the pairs sometimes can get reflected, but with a low probability of about $\lesssim 0.2$. To summarize, we can design using the driven impurity good quality pair filters which mainly leave through pairs of particles. 

In a similar fashion one can find configurations where the impurity acts as a single particle filter, i.e. blocks most pairs and only transmits single particles. This can in particular be realized in the limit of large interaction (cf.~$\vb{H}\upd{eff}$), a broad parameter regime exists where the single particle transmission is described by the low frequency expansion $\bar{T}\ind{s}^{\textrm{low}}$, 
whereas the pair transmission needs to be covered already in the high frequency limit and takes the form of the squared Bessel function.  Thus, in this parameter regime the single particle transmission is always much larger than the pair transmission. Using even a driving value $2 \lambda \approx 1.2$ leads to an almost vanishing pair transmission.

Tuning the amplitude of the driving $\lambda$ and the ratio of $U/\omega$ further situations can be realized as for example a blocking of both single particles and pairs. We summarize the discussed configurations in table \ref{tab:configurations}.  

Let us discuss in the following a few possible issues which might reduce the efficiency of the designed filters.
First, the effective model for the pair (\ref{equ:hamiltonian_pair}) is only applicable in the non-resonant case, where $U$ is not an integer multiple of $\omega$, because both $\gamma\ind{p}$ and $\mu\ind{p}$ diverge. In the resonant case, i.e.~$U/\omega$ is integer, a careful treatment of the pair breaking using the Lippmann-Schwinger equation shows that, as long as $J \ll \omega$, the resulting transmission is exactly given by the limiting expression of (\ref{equ:transmission_pair}). In that sense  Eq.~\ref{equ:transmission_pair} is valid for all parameters $U/\omega$.

\setlength\extrarowheight{3pt}
\begin{table}[hbtp]
\centering
\begin{tabular}{ccccc}
		\hline
		$|U|/\omega$ & $\lambda$ & $\bar{T}\ind{s}$ & $\bar{T}\ind{p}$ & Description\\
		\hline
		\multirow{2}{*}{1.46} & $0.9$ & $0.652$ & $< 10^{-3}$ & Single particle filter\\
		& $2.4$ & $< 10^{-3}$ & $0.794$ & Pair filter\\
		\hline
		\multirow{3}{*}{5.7}
		& $1.2$ & $0.45$ & $< 10^{-3}$ & Single particle filter\\
		& $2.64$ & $0.013$ & $0.006$ & Both blocked\\
		& $5.52$ & $< 10^{-3}$ & $0.81$ & Pair filter\\
		\hline
		\multirow{2}{*}{$\infty$} & $1.2$ & $0.45$ & $< 10^{-3}$ & Single particle filter\\
		& $2.64$ & $0.013$ & $0.007$ & Both blocked\\
		\hline
	\end{tabular}
\caption{Tunable impurity: A summary of parameters configurations yielding different filters.}
\label{tab:configurations}
\end{table}

\begin{figure}[hbtp]
\centering
\includegraphics[width=.43\textwidth]{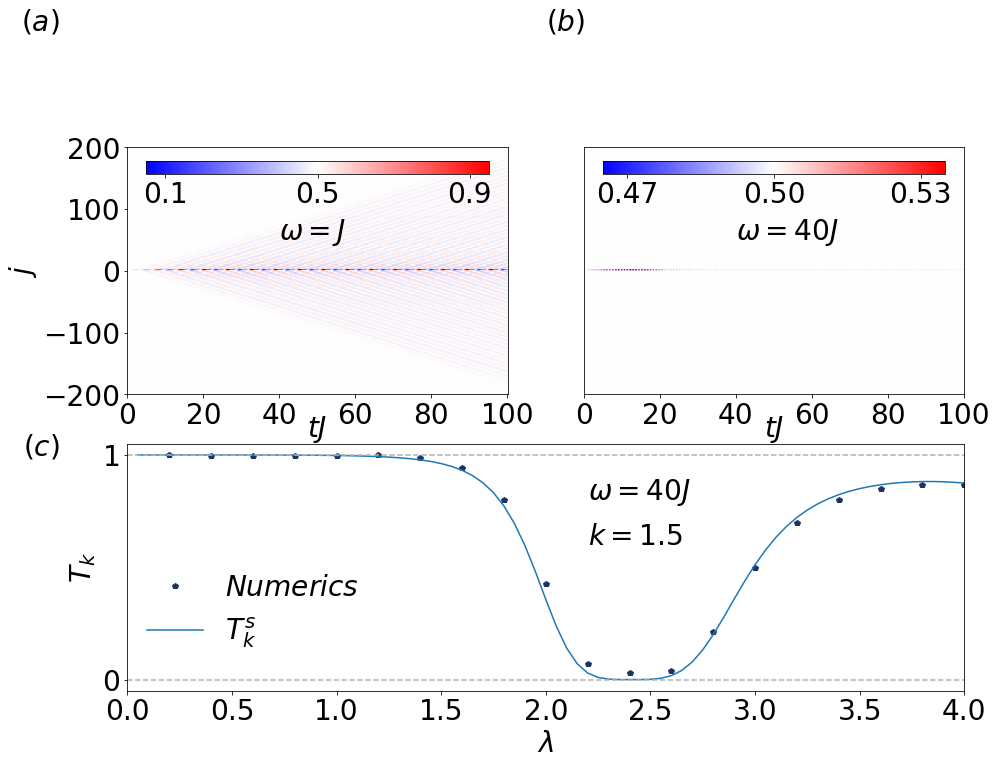}
\caption{ Time evolution of the density for a chain of $L=400$ sites with half filled non-interacting fermions with the driven impurity at the central site with frequency (a) $\omega=J$ and (b)$\omega=40J$ . The driving amplitude is ramped up linearly up to the maximum value of $\lambda=2.4$ in the ramp time $t_{ramp}=20/J$. (c) The dependence of the single particle transmission $T_k$ on the driving amplitude $\lambda$. 
  The driving frequency is $\omega=40J$ and the incoming wave is a wave packet of a single hole with unitless momentum $k=1.5$ and width $\sigma=50$ sites in position space. The analytical result $T_k^s$ (solid line) agrees well with the full numerical simulation (symbolds). }
\label{fig:background}
\end{figure}

Second, up to now we have focussed on the situation of an incoming single particle or of an incoming pair of two particles in an otherwise empty system. In order to verify the stability of our results, we now consider an incoming particle on top of a non-interacting background. In Fig.~\ref{fig:background} we show the density evolution in time for two different frequencies, $\omega=J$ which is within the bandwidth and $\omega=40J$ which is much larger than the other energy scales of the system. The results are obtained using a time-dependent exact diagonalization method for a system of $L=400$ sites. We find that the periodic driving of the impurity can induce considerable density oscillations in particular at low and intermediate driving frequencies. However, even though these oscillations are present, the transmission can be extracted by subtracting the densities of a system with and without incoming excitation. For very large frequencies the effect of driving on the background is mainly localized close to the impurity. As shown in Fig.~\ref{fig:background} the extracted transmission from the non-perturbed background agrees well with the expected transmission for a single particle $T_k^s$.
This means that the transmission through the driven impurity seems stable also in the presence of a background.

To summarize we have designed a quantum device which can act as a filter for pairs and single particles. It consists of a periodically driven impurity in the chain of interacting Fermions. Setting suitable driving parameters the driven impurity can be used to block incoming single particles or pairs. This in particular required taking carefully the limit of large frequency and interaction at the same time, since otherwise the interesting features do not occur. We quantified the quality of these filters and investigated their stability against a non-interacting background. One of the advantages of the Floquet-engineered impurity is that it does not only provide possible configurations for single particle and pair filters, but would also allow to quickly change between them in experiments by tuning, for example, the driving amplitude. Within this paper, we focussed on the momentum integrated transmission. Using the given equations, it can be easily used in order to design also momentum dependent filters.

\textit{Acknowledgments}
 We thank M.~K\"ohl, H.~Ott and I.~Schneider for stimulating discussions. We acknowledge funding from the Deutsche Forschungsgemeinschaft (DFG, German Research Foundation) in particular under project number 277625399 - TRR 185 (B3,B4) and project number 277146847 - CRC 1238 (C05) and under Germany’s Excellence Strategy – Cluster of Excellence Matter and Light for Quantum Computing (ML4Q) EXC 2004/1 – 390534769 and the European Research Council (ERC) under the Horizon 2020 research and innovation programme, grant agreement No.~648166 (Phonton).

\bibliographystyle{h-physrev}

\section{Appendix A: Gauge transformation}
\label{sec:app_gauge}
In this section we would like to explain how one derives a Fermi-Hubbard like Hamiltonian with time-periodic hopping from (\ref{equ:hamiltonian}) using the gauge transformation $\vb{U}(t) = e^{-i\lambda(\vb{n}_{0\uparrow} + \vb{n}_{0\downarrow})\sin{\omega t}}$. For such a time-dependent gauge transformation the resulting Hamiltonian is given by:
\begin{align}
\vb{H}\upd{g}(t) = \vb{U}(t)^\dagger\vb{H}(t)\vb{U}(t) - i\dv{t}\vb{U}(t)^\dagger \vb{U}(t)
\end{align}

The gauge transformation is defined as such that the last term exactly cancels the driving term in (\ref{equ:hamiltonian}). It remains to compute the first part. To this end observe that:

\begin{align}
\vb{U}(t)^\dagger \vb{c}_{n\sigma} \vb{U}(t) = \begin{cases}
	e^{-i\lambda\sin{\omega t}} \vb{c}_{0\sigma} & n = 0\\
	\vb{c}_{n\sigma} & \mathrm{else}. 
\end{cases}
\end{align}
By inserting $1 = \vb{U}(t) \vb{U}(t)^\dagger$ into Hamiltonian (\ref{equ:hamiltonian}) one can use this result to compute the gauge transformation of all other quantities. For example:
\begin{align}
\vb{U}(t)^\dagger\vb{c}_{n\sigma}^\dagger\vb{c}_{n+1\sigma}\vb{U}(t)\\ = \begin{cases} e^{-i\lambda\sin{\omega t}} \vb{c}_{n\sigma}^\dagger\vb{c}_{n+1\sigma} & n = -1\\
e^{i\lambda\sin{\omega t}} \vb{c}_{n\sigma}^\dagger\vb{c}_{n+1\sigma} & n = 0\\
\vb{c}_{n\sigma}^\dagger\vb{c}_{n+1\sigma} & \mathrm{else}
\end{cases}
\end{align}

while any $\vb{c}_{n\sigma}^\dagger\vb{c}_{n\sigma}$ is unaffected by the gauge transformation.

We denote the prefactors in front of the hopping terms as $g_n(\omega t)$. Collecting all terms finally leads to \begin{align}
 \vb{H}\upd{g}(t) = -J&\sum_{n\sigma} \qty(g_n(\omega t)\vb{c}_{n\sigma}^\dagger\vb{c}_{n+1\sigma} + \mathrm{h.c.}) + U\sum_n \vb{n}_{n\uparrow}\vb{n}_{n\downarrow} \label{equ:hamiltonian_gauge}
\end{align}
where $g_n(\phi) = 1$ except for $g_{-1}(\phi) = e^{-i\lambda \sin(\phi)}$ and $g_{0}(\phi) = e^{i\lambda \sin(\phi)}$.

\section{Appendix B: Calculation of the transmission in the effective model}
\label{sec:app_trans}
Consider a pair coming from the left with momentum $k$ and energy $\epsilon_k = -2J\ind{p}\cos{k}$. We would like to calculate the transmission amplitude. The wavefunction has the general form:
\begin{align}
\psi_n = \begin{cases} 
      e^{ikn} + r_k e^{-ikn} & n<0\\
      \psi_0 & n=0 \\
      t_k e^{ikn} & n>0
   \end{cases}.
   \label{equ:app_trans_wavefunction}
\end{align}
The Schr\"odinger equation evaluated at sites $n = -1,0$ and $1$ gives the following set of equations:
\begin{align}
\psi_{-2} + \gamma\ind{p}\psi_0 + \mu\ind{p} \psi_{-1} &= 2\cos(k) \psi_{-1}\\
\gamma\ind{p}\psi_{-1} + \gamma\ind{p}\psi_1 + 2\mu\ind{p} \psi_0 &= 2\cos(k) \psi_0\\
\gamma\ind{p} \psi_0 + \psi_2 + \mu\ind{p} \psi_1 &= 2\cos(k) \psi_1.
\end{align}

Inserting the wavefunction (\ref{equ:app_trans_wavefunction}) and solving the equations gives:
\begin{align}
t_k &= \frac{\gamma\ind{p}^2}{1-\mu\ind{p}e^{ik}}\frac{-2i\sin k}{\qty(2\cos k-2\mu\ind{p})\qty(1-\mu\ind{p} e^{ik})-2\gamma\ind{p}^2e^{ik}}.
\end{align}
In order to obtain the transmission probability we have to square this expression $T_k = |t_k|^2$ which gives (\ref{equ:transmission_pair}).

\end{document}